\IfFileExists{elsarticle.cls}{%
  \documentclass[preprint,12pt,a4paper]{elsarticle}%
  \biboptions{numbers,sort&compress}%
}{%
  \documentclass[11pt,a4paper]{article}%
  \usepackage{elsfallback}%
}

\usepackage{graphicx}
\usepackage{amsmath,amssymb}
\usepackage{booktabs}
\usepackage{xcolor}
\usepackage{url}
\usepackage{array}
\usepackage{listings}
\usepackage[section,above,below]{placeins}
\usepackage[hidelinks]{hyperref}

\newcommand{\code}[1]{\texttt{#1}}
\newcommand{\hd}[1]{\begin{tabular}[b]{@{}c@{}}#1\end{tabular}}
\journal{Computer Physics Communications}

\begin{document}

\begin{frontmatter}

\title{SKFitter: symmetry-resolved Slater--Koster tight-binding models from first principles for s, p, d and f electrons, spin--orbit coupling and magnetism}

\author[iitb]{Sundeep Kapila\corref{cor1}}
\ead{sundeep.kapila@swasth.org}
\author[iitb]{Bhaskaran Muralidharan}
\author[ikst]{Satadeep Bhattacharjee\corref{cor1}}
\ead{s.bhattacharjee@ikst.res.in}
\cortext[cor1]{Corresponding authors}
\affiliation[iitb]{organization={Department of Electrical Engineering, Indian Institute of Technology Bombay},
  city={Mumbai}, postcode={400076}, country={India}}
\affiliation[ikst]{organization={Indo-Korea Science and Technology Center}, city={Bengaluru}, country={India}}

\begin{abstract}
Constructing accurate and physically interpretable tight-binding Hamiltonians from first-principles calculations remains challenging, particularly for materials exhibiting complex orbital hybridization, spin–orbit coupling, and magnetic order. We present SKFitter, an open-source Python framework for constructing and optimizing symmetry-resolved Slater–Koster tight-binding models directly from density functional theory (DFT) calculations performed using Quantum ESPRESSO or VASP. The framework supports $s$, $p$, $d$, and $f$ orbitals, incorporates on-site spin–orbit coupling and site-resolved magnetic exchange interactions, and accommodates both orthogonal and non-orthogonal basis representations. A central feature is the symmetry-resolved parameterization of orbital manifolds, which allows independent two-centre hopping integrals between symmetry-allowed orbital groups and across distinct neighbour shells. Crystal-symmetry constraints are enforced during model construction, eliminating symmetry-forbidden parameters and preventing incompatible orbital decompositions. SKFitter additionally employs DFT orbital projections to identify missing orbital contributions and guide systematic basis-set refinement. Model parameters are optimized through weighted band-structure fitting, and the resulting Hamiltonian can be used to evaluate Berry curvature, anomalous and spin Hall conductivities, and topological invariants. We demonstrate the framework using eight representative materials spanning semimetals, semiconductors, transition metals, $f$-electron compounds, and magnetic systems. Symmetry-resolved models achieve band-energy errors of 9–41 meV within $\pm 1$ eV of the Fermi level. The fitted Hamiltonians reproduce characteristic features of the spin Hall response of Pt, the spin–orbit-coupling dependence of the anomalous Hall conductivity of Fe, and the momentum-dependent altermagnetic spin splitting of MnTe without introducing an altermagnet-specific interaction. For EuMn$_2$Bi$_2$, projection-guided orbital refinement reduces the near-Fermi-level fitting error from 120 to 32 meV. Together, these results establish SKFitter as a flexible and reproducible framework for developing symmetry-consistent, physically interpretable tight-binding models of complex electronic and magnetic materials.
\end{abstract}

\begin{keyword}
tight binding \sep Slater--Koster \sep density functional theory \sep spin--orbit coupling \sep altermagnetism \sep Berry curvature
\end{keyword}

\end{frontmatter}

\section*{Program summary}
\noindent
\textit{Program title:} SKFitter\\[2pt]
\textit{CPC Library link to program files:} (to be added by Technical Editor)\\[2pt]
\textit{Developer's repository link:} \url{https://github.com/sundeep-kapila/skfitter}; release archive: doi:10.5281/zenodo.23719846\\[2pt]
\textit{Licensing provisions:} GNU Affero General Public License v3.0 (AGPL-3.0-only); example and test data under CC BY 4.0\\[2pt]
\textit{Programming language:} Python ($\geq3.10$)\\[2pt]
\textit{Supplementary material:} none\\[2pt]
\textit{External routines/libraries:} NumPy~\cite{Harris2020}, SciPy~\cite{Virtanen2020}, Matplotlib~\cite{Hunter2007}, pymatgen~\cite{Ong2013} (spglib~\cite{Togo2024}); tomli on Python 3.10\\[2pt]
\textit{Nature of problem:} Tight-binding Hamiltonians are the working models of band topology, transport and linear response. Slater--Koster (SK) models have few, interpretable parameters, but building one that is accurate for a given material is largely manual: published parameter sets rarely cover f electrons, spin--orbit coupling (SOC) and magnetic order together; one integral per orbital family and bond type cannot express crystal-field splittings; and antiferromagnetic or altermagnetic order must be put in by hand.\\[2pt]
\textit{Solution method:} SKFitter reads the bands, k-path, Fermi level, electron count, magnetic moments and orbital projections of a Quantum ESPRESSO or VASP run, builds an SK Hamiltonian for chosen s, p, d and f orbitals (whole families or symmetry-resolved groups, with two-centre integrals for every pair of element and orbital group), on-site SOC and exchange along each site's moment direction, and fits its parameters to the DFT bands by bounded least squares with Fermi-level and high-symmetry-point weights and warm starts from simpler models. Integrals that vanish by symmetry on every bond are pruned and splits that the site symmetry does not allow are refused. The orbitals a model lacks are ranked from the DFT projections, with the groups the symmetry allows for each. The fitted model gives band structures, the orbital character of the DFT states, Berry and spin Berry curvature, anomalous and spin Hall conductivities, Wilson loops, Chern numbers, hybrid Wannier charge centres and Z2 indices.\\[2pt]
\textit{Additional comments including restrictions and unusual features:} Command-line tool (\code{skfitter}) driven by a TOML configuration, and a Python package. Runs on Linux, macOS and Windows. A fit takes seconds to minutes for small cells and up to hours for large magnetic f--d--p models (EuMn$_2$Bi$_2$: 216 parameters, 141 k-points $\times$ 78 bands); \code{skfitter check} evaluates published parameters in seconds. The distribution includes the Quantum ESPRESSO data of eight benchmark systems, their configurations and reference results, and scripts that regenerate the tables and figures of this paper.

\section*{Highlights}
\begin{itemize}
\item Slater--Koster models fitted to Quantum ESPRESSO or VASP bands, s to f orbitals
\item Symmetry-resolved orbital groups and integrals for every element pair
\item Missing orbitals ranked from DFT projections, with symmetry-allowed groups
\item Spin--orbit coupling and exchange along DFT moment directions, incl.\ altermagnets
\item Berry curvature and Hall conductivities; eight benchmarks reproducible from data
\end{itemize}

\section{Introduction}\label{sec:intro}
\begin{figure}[!tbp]
\centering
\includegraphics[width=\textwidth]{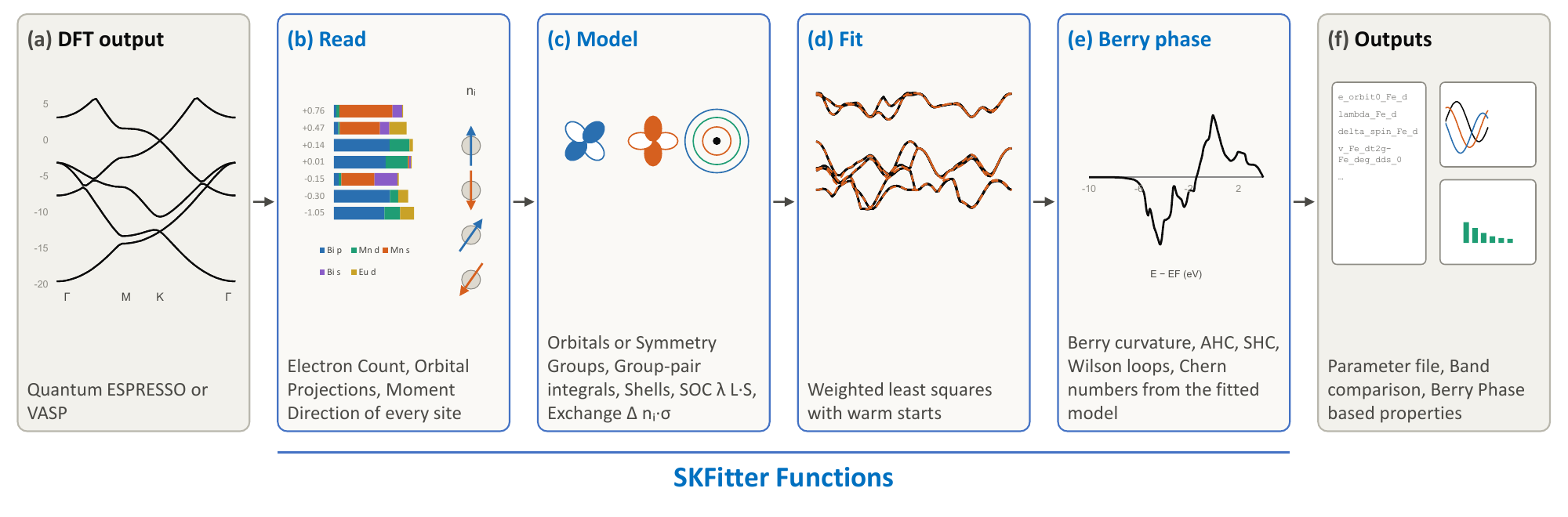}
\caption{\textbf{SKFitter workflow.} (a) Input: the band-structure output of Quantum ESPRESSO or VASP (graphene bands). (b) Reading the bands, the electron count, the orbital projections (EuMn$_2$Bi$_2$ states at $\Gamma$) and the moment direction $\mathbf n_i$ of every site. (c) Model: orbitals or symmetry groups, group-pair integrals, neighbour shells, SOC $\lambda\,\mathbf L\cdot\mathbf S$ and exchange $\Delta\,\mathbf n_i\cdot\boldsymbol\sigma$. (d) Fit: weighted least squares (Eq.~(\ref{eq:loss})) with warm starts (2H-MoS$_2$: DFT solid, fit dashed). (e) Berry-phase properties of the fitted model: Berry curvature, anomalous and spin Hall conductivities, Wilson loops and Chern numbers (Pt spin Hall conductivity). (f) Outputs: parameter file, band comparison and Berry-phase properties. Steps (b)--(e) are functions of SKFitter, run as commands of the \code{skfitter} program (Table~\ref{tab:commands}).}
\label{fig:workflow}
\end{figure}

Tight-binding Hamiltonians are the common language of band topology, quantum transport and linear response~\cite{Xiao2010}. They make Brillouin-zone integrals cheap, scale to devices and heterostructures, and their parameters carry chemical meaning. Two routes dominate. Maximally localised Wannier functions reproduce DFT bands closely but give dense, basis-dependent Hamiltonians whose parameters are hard to interpret or transfer~\cite{MarzariVanderbilt1997,Marzari2012,Pizzi2020,Gresch2018}. Slater--Koster (SK) models give a few two-centre integrals with clear meaning~\cite{SlaterKoster1954,Harrison1980}, but published parameter sets are tied to particular materials and rarely cover f orbitals, spin--orbit coupling (SOC) and magnetism together~\cite{Papaconstantopoulos2015,Elstner1998}.

Fitting SK models to DFT bands is not new, but most implementations stop at s, p and d orbitals, use one integral per orbital family and bond type, and treat magnetism as a rigid shift. Three gaps follow. First, f electrons with SOC: rare-earth and actinide compounds need f orbitals, their two-centre integrals ($ff\sigma$ to $ff\phi$, $df$, $pf$)~\cite{Takegahara1980,PodolskiyVogl2004} and on-site SOC in the same model. Second, crystal-field structure: in a crystal the d (or p, f) orbitals split into symmetry groups whose hoppings and exchange differ; one integral per family cannot express that. Third, magnetic order: antiferromagnets and altermagnets need site-resolved exchange; their key physics, such as the spin splitting of MnTe~\cite{Smejkal2022PRX031042,Smejkal2022PRX040501}, lives in how exchange and hopping combine.

SKFitter closes these gaps with a general SK construction for s, p, d and f orbitals, symmetry-resolved orbital groups with group-pair integrals and per-group exchange, integrals for every element pair and, optionally, every neighbour shell, and moment directions read from the DFT run (Fig.~\ref{fig:workflow}). Which orbitals a model needs is decided from the DFT orbital projections: the program ranks the orbitals the model lacks and names the groups the crystal symmetry allows for each. This paper describes the method (Section~\ref{sec:method}), the program, its input and its use (Section~\ref{sec:program}), and test runs on eight materials whose data and configurations are distributed with the code (Section~\ref{sec:tests}): the benchmark of the model classes, the Berry-phase properties of Fe and Pt against first-principles references, the altermagnetic splitting of MnTe and the construction of a model for EuMn$_2$Bi$_2$ from the orbitals the projections suggest. Section~\ref{sec:compare} compares SKFitter with other SK codes.

\section{Method}\label{sec:method}

\subsection{Hamiltonian}\label{sec:hamiltonian}
The Hamiltonian is a sum of on-site energies, two-centre SK hoppings, on-site SOC and a site-resolved exchange splitting, built for any combination of s, p, d and f orbitals:
\begin{equation}
\begin{split}
H(\mathbf k)={}&\sum_{i\alpha}\varepsilon_{i\alpha}\,c^\dagger_{i\alpha}c_{i\alpha}
+\sum_{ij,\mathbf R}e^{i\mathbf k\cdot\mathbf R}\,t_{i\alpha,j\beta}(\mathbf d)\,c^\dagger_{i\alpha}c_{j\beta}\\
&+\sum_i\lambda_{i\ell}\,\mathbf L\cdot\mathbf S
-\frac12\sum_{i\alpha}\Delta_{i\alpha}\,\mathbf n_i\cdot c^\dagger_{i\alpha}\boldsymbol\sigma\, c_{i\alpha} .
\end{split}
\label{eq:H}
\end{equation}
For a bond direction $\mathbf d$, the hopping block between orbital families $\ell$ and $\ell'$ is a sum over the bond-axis quantum number $m$ of an integral $V_{\ell\ell'm}$ times a unit block $U_m(\mathbf d)$ obtained by rotating real spherical harmonics. The same code path covers $ss\sigma$ through $ff\phi$, including the $df$ and $pf$ integrals needed for rare-earth compounds. Every integral belongs to one pair of elements and orbitals: bonds of different element pairs never share an integral, even at the same distance (Mn--Mn and Te--Te bonds of length $a$ in MnTe), and an $sp\sigma$ integral depends on which element carries the s orbital. Overlaps can be added with the same construction (non-orthogonal models, generalized eigenvalue problem); property calculations then use the L\"owdin-orthogonalised Hamiltonian $S^{-1/2}HS^{-1/2}$~\cite{Lowdin1950}.

\textbf{Symmetry-resolved orbital groups.} A family can be split into groups that the crystal's point group leaves invariant: $p\to p_z, p_{xy}$; $d\to t_{2g}, e_g$ (cubic) or $A_1$ ($d_{z^2}$), $E_1$, $E_2$ (hexagonal and trigonal); $f\to a_{2u}$ ($xyz$), $t_{1u}$, $t_{2u}$ (cubic). Each group has its own on-site energy and, if magnetic, its own exchange splitting. Each pair of groups $A,B$ has its own integrals,
\begin{equation}
H_{AB}(\mathbf R)=\sum_m V^{AB}_m\,U_m(\mathbf d)\big|_{A,B},
\label{eq:group}
\end{equation}
where $U_m|_{A,B}$ is the block of the unit matrix between the orbitals of $A$ and $B$. The cubic f groups are combinations rather than subsets of the real f orbitals; with $B_A$ the rows of group $A$ in that basis, the block is $B_A U_m B_B^{\mathsf T}$, and the SOC matrices are restricted the same way. Integrals that vanish on every bond of the structure are not fitted: on the Mn--Mn bond along $c$ in MnTe only $d_{z^2}\sigma$, $E_1\pi$ and $E_2\delta$ survive. The program checks that every group is invariant under the crystal's symmetry operations (from spglib~\cite{Togo2024} via pymatgen~\cite{Ong2013}) and refuses a split that the structure does not allow. Optionally, on-site couplings between groups of a site that its symmetry allows are added as parameters.

\textbf{Distance dependence.} The neighbour shells are counted for each element pair, shell 0 being the pair's nearest bond. Integrals follow either a power law $V(r)=V_0\,(d_0/r)^{\eta}$, with one $V_0$ and $\eta$ per integral and $d_0$ the nearest bond of the element pair, or take independent values $V_n$ in each shell $n$ of the pair (``per shell''); bonds slightly off their shell distance are scaled as $V_n(r_n/r)^\eta$. Parameters are named after what they describe: \code{v\_Te\_p-Mn\_d\_pds\_1} is the $pd\sigma$ integral between Te p and Mn d on the second Mn--Te shell, \code{v\_Bi\_pz-Mn\_dz2\_pds\_0} its counterpart between two orbital groups, \code{eta\_\ldots} a decay exponent and \code{s\_\ldots} an overlap.

\textbf{Spin--orbit coupling and exchange.} $\lambda_\ell\,\mathbf L\cdot\mathbf S$ acts on each p, d and f family, or on the families chosen in the configuration, in the on-site form of Chadi~\cite{Chadi1977}. In the last term of Eq.~(\ref{eq:H}), $c_{i\alpha}=(c_{i\alpha\uparrow},c_{i\alpha\downarrow})^{\mathsf T}$, $\boldsymbol\sigma$ are the Pauli matrices and $\mathbf n_i$ is the unit vector along the moment of site $i$: the exchange splitting $\Delta_{i\alpha}$ (one per family or symmetry group) lowers the states with spin along $\mathbf n_i$ by $\Delta_{i\alpha}/2$ and raises the opposite ones. The directions are read from the DFT run: the per-site magnetization of a non-collinear pw.x run or the MAGMOM vectors of VASP, and $\mathbf n_i=s_i\hat{\mathbf z}$ with $s_i=\pm1$ for a collinear run. Antiferromagnets, altermagnets and non-collinear orders therefore keep their arrangement, and with SOC the moment direction relative to the lattice enters the model.

\subsection{Fitting}\label{sec:fitting}
The parameters $\theta$ (on-site energies, integrals, $\lambda$, $\Delta$) minimise the weighted mean squared band error
\begin{equation}
L(\theta)=\frac{1}{N}\sum_{\mathbf k}\sum_{n}w_{n\mathbf k}\,\big[E^{\rm TB}_{n\mathbf k}(\theta)-E^{\rm DFT}_{n\mathbf k}\big]^2
\label{eq:loss}
\end{equation}
over the $N$ pairs $(n,\mathbf k)$ of the fit window: the DFT bands from $n_{\rm below}$ bands below to $n_{\rm above}$ bands above the last occupied one, at every k-point of the band path. Energies are relative to $E_F$. The TB eigenvalues are sorted at each $\mathbf k$ and matched to the DFT bands by their position relative to the last occupied band, which is set by the electron count of the model's orbitals. The weights are a product of optional factors, all equal to 1 by default:
\begin{equation}
w_{n\mathbf k}=\Big[1+(w_F-1)\,e^{-(E^{\rm DFT}_{n\mathbf k}/\varepsilon_F)^2}\Big]\Big[1+(w_S-1)\,p_{\mathbf k}\Big]\,m_{n\mathbf k}.
\label{eq:weights}
\end{equation}
The first factor emphasises the bands within about $\varepsilon_F$ of $E_F$, which control Berry-curvature and transport properties. The second emphasises the high-symmetry points: $p_{\mathbf k}=1$ at the ends of every path segment and falls linearly to 0 over $n_S$ neighbouring k-points. The mask $m_{n\mathbf k}\in\{0,1\}$ optionally drops DFT states whose projections on the model's orbitals sum to less than 0.5 (for example vacuum states of a layer).

The minimisation uses SciPy's~\cite{Virtanen2020} bounded trust-region reflective least squares~\cite{Branch1999} on the residuals $\sqrt{w_{n\mathbf k}}\,[E^{\rm TB}_{n\mathbf k}-E^{\rm DFT}_{n\mathbf k}]$, with bounds on every parameter and one or more starts (the first from the starting values, the others perturbed); differential evolution~\cite{StornPrice1997} is available as a global alternative. The finite-difference Jacobian can be evaluated on several threads (\code{workers}). The band evaluation is vectorised over k-points with precomputed bond phases and unit blocks and checked against a reference implementation. Starting values come from bundled parameter sets (Harrison's universal coefficients~\cite{Harrison1980} with NIST atomic energies~\cite{Kotochigova1997}, or DFTB sets~\cite{Elstner1998}), from the DFT band centres of each orbital, or from an earlier fit (a warm start): a decay-law fit starts a per-shell model, and a group integral starts at the family value it splits, so each refinement begins from the previous model's bands. Reported errors are unweighted RMSE, $\big(\frac1N\sum[E^{\rm TB}-E^{\rm DFT}]^2\big)^{1/2}$, over all fitted bands or over the DFT states within $E_F\pm1$~eV.

\subsection{Choosing the orbitals}\label{sec:suggest}
\code{skfitter character -{}-suggest} ranks the orbital families a model lacks from the DFT orbital projections over all k-points of the fit window. Each state carries the Fermi-level factor of Eq.~(\ref{eq:weights}), its projections are summed per element and family (a channel such as Eu d), and for every channel the program reports: the largest weight in any state (\emph{absolute}); the largest weighted average over $\mathbf k$ of any band (\emph{weighted}); its share of the weight the model's orbitals cannot carry (\emph{missing share}), that is, what adding it can explain; its projection-weighted mean energy, a starting on-site energy; its occupied weight as a fraction of the capacity $2(2\ell+1)$ per site (\emph{filled}), which decides how many electrons it adds to the model; and the number of parameters it adds. A channel is suggested when its absolute or weighted value reaches a cutoff (0.1 by default), and the command prints the orbital line of the extended model; the configuration is not changed. Projections never sum to one, and the weight on no atomic orbital is reported as a bound on what added orbitals can explain. For every p, d and f shell the table also gives the finest split into SKFitter's orbital groups that the crystal symmetry allows on all sites of the element, the number of on-site crystal-field or mixing terms the site symmetry still allows inside the shell, and the site point group; for MnTe it gives $A_1/E_1/E_2$ for Mn d (with one further on-site term on the $\bar3m$ site) and $p_z/p_{xy}$ for Te p, the split of the model of Section~\ref{sec:mnte}.

\subsection{Berry-phase properties}\label{sec:berrymethod}
The Berry curvature~\cite{Berry1984,Xiao2010} of the fitted model and the intrinsic Hall conductivity are computed from the Kubo formula~\cite{Kubo1957,TKNN1982},
\begin{gather}
\Omega_n^z(\mathbf k)=-2\,\mathrm{Im}\sum_{m\neq n}\frac{\langle n|\partial_{k_x}H|m\rangle\langle m|\partial_{k_y}H|n\rangle}{(E_n-E_m)^2},\\
\sigma_{xy}=-\frac{e^2}{\hbar}\frac{1}{VN_k}\sum_{\mathbf k,n}f_n\,\Omega_n^z(\mathbf k),
\end{gather}
on a $\Gamma$-centred mesh with adaptive refinement of the k-points with large curvature~\cite{Yao2004,Wang2006}. The spin Hall conductivity uses the conventional spin current $\tfrac{\hbar}{4}\{\sigma_z,\mathbf v\}$~\cite{Sinova2004,Guo2008,Qiao2018}. Wilson loops (Berry phases of the occupied bands on circles) around the valleys $K$ and $K'$~\cite{Yu2011,Xiao2007} and Chern and valley Chern numbers from the link variables of a plaquette mesh~\cite{Fukui2005} are available for two-dimensional hexagonal zones. The exchange term lowers the states with spin along $\mathbf n_i$; with the DFT majority spin along $+z$ (magnetization along $-z$), $\sigma_{xy}$ of Fe is negative in this convention and is plotted as $-\sigma_{xy}$ in Fig.~\ref{fig:berry}. Hybrid Wannier charge centres from Wilson loops along a reciprocal lattice vector give the Chern number of a plane and, with time-reversal symmetry, its $Z_2$ index~\cite{Yu2011,Vanderbilt2011}.

\section{Program description}\label{sec:program}

\subsection{Structure}
SKFitter is a Python package (\code{skfitter}) with a command-line program of the same name (Table~\ref{tab:modules}). The numerics are in \code{skfitter.core}: the model, the vectorised Hamiltonian, the fit and the Berry-phase routines. \code{skfitter.workflow} exposes the steps of a calculation as functions (reading and caching DFT data, building and fitting a model, property runs, plots), on which the command line is built; graphical front ends can use the same functions. Fitted models are saved as parameter files and can be reloaded as model objects (\code{skfitter.model}) that offer a common tight-binding interface to the property calculators (\code{skfitter.properties}).

\begin{table}[!tbp]
\caption{Modules of the \code{skfitter} package.}
\label{tab:modules}
\small
\begin{tabular}{@{}>{\ttfamily}p{0.17\textwidth}p{\dimexpr0.83\textwidth-2\tabcolsep\relax}@{}}
\toprule
\textrm{Module} & Content \\
\midrule
core & SK Hamiltonian (s--f, groups, overlaps, SOC, exchange), DFT readers (QE XML, projwfc.x, VASP), fast band model, least-squares and differential-evolution fits, Berry-phase routines \\
workflow & project folders, DFT data cache, starting values and bounds, fit and property jobs, figures \\
structure & crystal, neighbour shells per element pair and the integrals each bond uses (for inspection and tests) \\
params & bundled parameter sets (Harrison, NIST, DFTB) and their mapping onto a model \\
model & model classes (\code{SKModel}, \code{TBModel}), common interface, file formats, field and strain modifiers \\
properties & Berry curvature, conductivities, densities of states on any model of \code{skfitter.model} \\
cli & the \code{skfitter} command \\
\bottomrule
\end{tabular}
\end{table}

\subsection{Installation and tests}
SKFitter needs Python 3.10 or newer; NumPy, SciPy, Matplotlib and pymatgen are installed with it:
\begin{lstlisting}
git clone https://github.com/sundeep-kapila/skfitter.git
cd skfitter
pip install ".[test]"
pytest -q
\end{lstlisting}
The test suite (230 tests, about three minutes) runs on Quantum ESPRESSO data shipped with the code and covers the SK blocks of every orbital pair, the symmetry check and pruning of the groups, the element-pair integrals and the reading of parameter files with the earlier element-independent names, the fast band model against the reference implementation, the readers, the fits, the moment directions and the Berry-phase routines against independent implementations. It runs on Linux, macOS and Windows in continuous integration.

\subsection{Input}
A calculation needs the output of a band-structure run along a high-symmetry path: from Quantum ESPRESSO~\cite{Giannozzi2009,Giannozzi2017} the XML file of the non-self-consistent bands run, its pw.x input (k-path labels), optionally the SCF output (Fermi level, moments) and the projwfc.x projections; from VASP~\cite{KresseFurthmuller1996} vasprun.xml and the line-mode KPOINTS file. A TOML file names these files and describes the model, the fit and the property calculations. For 2H-MoS$_2$ (Section~\ref{sec:tests}; \code{examples/mos2/skfitter.toml}, abridged):
\begin{lstlisting}
name = "mos2"
output = "out"

[dft]
code = "espresso"
xml = "data/qe_bands.xml.gz"
bands_in = "data/qe_bands.in"
scf = "data/qe_scf.out"
projwfc = "data/qe_projwfc_up.gz"

[model]
orbitals = { Mo = ["dz2", "d_e2", "d_e1"], S = ["pxy", "pz"] }
bands_below = 4
bands_above = 2
neighbours = 9
hopping = "shell"
soc = true
electrons = 28

[fit]
starts = 2
fermi_weight = 60
fermi_window = 0.7
start_from = "reference/mos2_fit.json"
\end{lstlisting}
A property calculation is one \code{[[property]]} table, here the Pt spin Hall conductivity of Fig.~\ref{fig:berry}(c) (\code{examples/platinum/skfitter.toml}):
\begin{lstlisting}
[[property]]
type = "shc"
nk = [80, 80, 80]
emin = -10.0
emax = 4.0
num_efermi = 281
\end{lstlisting}
In a configuration, \code{[model]} sets the orbitals per element (families \code{s, p, d, f} or groups such as \code{d\_t2g}, \code{dz2}, \code{pxy}, \code{f\_a2u}), the magnetic orbitals, the families with SOC, the fit window, the number of neighbour shells, the distance law, SOC, overlaps and the electron count. \code{[fit]} sets the method, the number of starts, the weights of Eq.~(\ref{eq:weights}), the starting values (\code{start\_set} or \code{start\_from}) and optional initial values, bounds and fixed parameters. Each \code{[[property]]} table is one calculation (\code{berry\_path}, \code{berry\_map}, \code{wilson}, \code{chern}, \code{spin\_berry\_path}, \code{ahc}, \code{shc}, \code{shc\_temperature}, \code{wcc}) and may change parameters of the fitted model for that calculation only, as in the spin--orbit scan of Fig.~\ref{fig:berry}(b). The manual (\code{docs/manual.md}) lists every key.

\subsection{Commands and output}
Table~\ref{tab:commands} lists the commands. \code{skfitter info} shows the DFT data, the crystal, the neighbour shells of every element pair and the parameters of a model before it is fitted; \code{skfitter character} gives the orbital character of the DFT states at a k-point and, with \code{-{}-suggest}, the ranking of Section~\ref{sec:suggest} (used in Section~\ref{sec:eu}); \code{skfitter fit} writes the fitted parameters (\code{<name>\_fit.json} and a text version), the DFT and tight-binding bands (\code{<name>\_bands.dat} and a plot) and an error report (\code{<name>\_report.json}); \code{skfitter berry} writes one data file and one plot per property. All output files are plain text or JSON.

\begin{table}[!tbp]
\caption{Commands of the \code{skfitter} program (\code{skfitter COMMAND CONFIG}).}
\label{tab:commands}
\small
\begin{tabular}{@{}>{\ttfamily}p{0.14\textwidth}p{\dimexpr0.86\textwidth-2\tabcolsep\relax}@{}}
\toprule
\textrm{Command} & Action \\
\midrule
info & DFT data, crystal, neighbour shells per element pair, model size; \code{-{}-params}: parameters with starting values and bounds \\
character & orbital character of the DFT bands at a k-point (\code{-{}-k}: label or fractional coordinates); \code{-{}-suggest}: the orbitals the model lacks, ranked over all k-points, with their symmetry-allowed groups \\
check & evaluate the parameters of \code{start\_from} without fitting; compare with the reference report \\
fit & fit the model; \code{-{}-workers N} threads for the Jacobian \\
bands & bands of the saved fit; \code{-{}-spin}: spin-resolved bands of a collinear magnet \\
berry & the \code{[[property]]} calculations on the saved fit (\code{-{}-only}, \code{-{}-index}) \\
run & fit, then every property \\
params & the bundled parameter sets \\
\bottomrule
\end{tabular}
\end{table}

\subsection{Reproducing this paper}
The repository holds, for each of the eight systems of Section~\ref{sec:tests}, the Quantum ESPRESSO data (\code{examples/<id>/data}, settings in Table~\ref{tab:dft}), the configuration of the model of Table~\ref{tab:benchmark} (\code{skfitter.toml}), the configurations of the simpler models and of the EuMn$_2$Bi$_2$ build-up (\code{steps/}), and the fitted parameters with their error reports (\code{reference/}). Three levels of test run are provided:
\begin{lstlisting}
python paper/run_examples.py --only check  # published parameters: minutes
python paper/run_examples.py --jobs 12     # every fit and property: ~1 day
python paper/run_examples.py --summary     # the errors of every fit
python paper/make_figures.py               # figures and tables
\end{lstlisting}
\code{check} reproduces the errors of Table~\ref{tab:benchmark} exactly from the published parameters. The full run repeats every fit (warm-started as in the paper), the decay-law and per-shell models from the Harrison starting values, the orbital ranking and build-up of EuMn$_2$Bi$_2$ and the Fe and Pt conductivities; the runner starts the longest chains first, starts each calculation when the fits it starts from are done and skips finished ones. Most of the time goes into the EuMn$_2$Bi$_2$ models. Table~\ref{tab:timing} gives the size and running time of the final fits.

\begin{table}[!tbp]
\caption{Size and running time of the final fits of Table~\ref{tab:benchmark} (\code{skfitter fit}, one core per fit with up to 12 fits in parallel on a laptop with an Intel Core Ultra 7 165H processor (16 cores, 22 threads) and 64~GB of memory; started from the parameters of an earlier fit). \code{skfitter check} takes 2--60~s for every system. $^{a}$Two threads for the finite-difference Jacobian (\code{-{}-workers 2}), five EuMn$_2$Bi$_2$ fits in parallel with the other systems.}
\label{tab:timing}
\small
\begin{tabular}{@{}lrrrr@{}}
\toprule
System & Parameters & k-points & Bands & Time \\
\midrule
Graphene & 27 & 113 & 10 & 1.3 min \\
2H-MoS$_2$ & 87 & 211 & 12 & 69 min \\
Pt & 45 & 201 & 10 & 55 s \\
Fe & 72 & 122 & 10 & 2.4 min \\
$\gamma$-Ce & 125 & 181 & 10 & 2.0 min \\
CeO$_2$ & 63 & 181 & 13 & 1.6 min \\
MnTe & 63 & 121 & 30 & 13 min \\
EuMn$_2$Bi$_2$ & 216 & 141 & 78 & 11.9 h$^{a}$ \\
\bottomrule
\end{tabular}
\end{table}

\section{Test runs}\label{sec:tests}

The test set spans an sp$^2$ semimetal (graphene), a layered semiconductor (2H-MoS$_2$), two transition metals with strong SOC or ferromagnetism (Pt, Fe), two f-electron systems ($\gamma$-Ce, CeO$_2$), a collinear altermagnet (MnTe) and a magnetic f--d--p compound with SOC (EuMn$_2$Bi$_2$). All bands come from Quantum ESPRESSO 7.5 with the PBE functional (Table~\ref{tab:dft}). Every number in this section is produced by the configurations in the repository.

\subsection{Model classes across eight materials}\label{sec:benchmark}

\begin{figure}[!tbp]
\centering
\includegraphics[width=\textwidth]{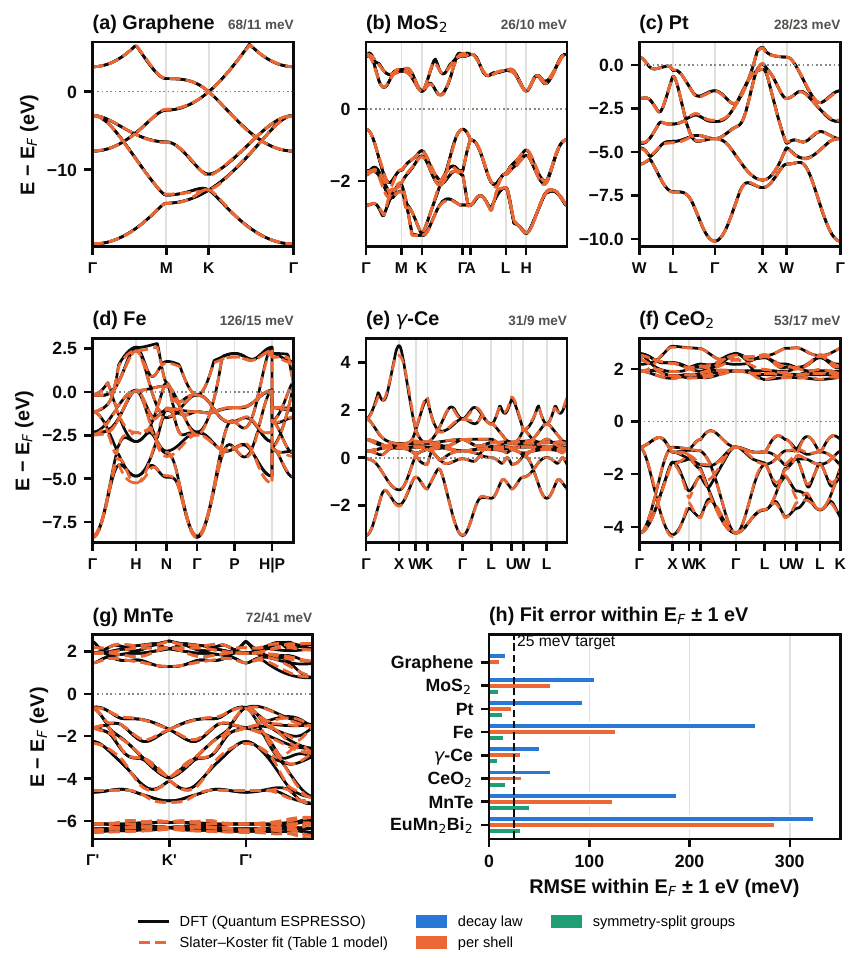}
\caption{\textbf{DFT and fitted Slater--Koster bands.} (a) Graphene, (b) 2H-MoS$_2$, (c) Pt, (d) bcc Fe, (e) $\gamma$-Ce, (f) CeO$_2$, (g) MnTe ($k_z=0.3$ path): DFT (solid black) and the SK model of Table~\ref{tab:benchmark} (dashed), with its RMSE over all bands / within $E_F\pm1$~eV. (h) RMSE within $E_F\pm1$~eV of the three model classes; for EuMn$_2$Bi$_2$ (bands in Fig.~\ref{fig:eu}) the split bar is the final model. The dashed line marks 25~meV.}
\label{fig:benchmark}
\end{figure}

For each system three model classes were fitted: one decay law per integral type and per-shell integrals, both with whole orbital families and fitted in a chain from the Harrison starting values (8 least-squares starts; 2 for EuMn$_2$Bi$_2$), and symmetry-split groups (Table~\ref{tab:benchmark}, Fig.~\ref{fig:benchmark}). In all of them the integrals are resolved by element pair (Section~\ref{sec:hamiltonian}).

\begin{table}[!tbp]
\caption{Benchmark models and errors. RMSE over all fitted bands and within $E_F\pm1$~eV for the model shown in Fig.~\ref{fig:benchmark}, and the $\pm1$~eV error of the three model classes (decay law, per shell, symmetry-split groups). Panel letters refer to Fig.~\ref{fig:benchmark}; the EuMn$_2$Bi$_2$ bands are shown in Fig.~\ref{fig:eu}, its split column is the final model with the extra Eu 5d, Mn 4s, Eu 6s and Bi 6s orbitals (Table~\ref{tab:suggest}).}
\label{tab:benchmark}
\scriptsize\setlength{\tabcolsep}{2pt}
\begin{tabular*}{\textwidth}{@{\extracolsep\fill}lp{3.2cm}cccccccc}
\toprule
 & & & & \multicolumn{2}{c}{model shown (meV)} & \multicolumn{3}{c}{$\pm1$ eV by class (meV)} \\
\cmidrule{5-6}\cmidrule{7-9}
Material & Orbitals and terms & Shells & Param. & all & $\pm1$ eV & decay & shell & split \\
\midrule
(a) Graphene & C s p, SOC, non-orthogonal & 3 & 27 & 68 & 11 & 17 & 11 & -- \\
(b) MoS$_2$ & Mo d ($A_1$, $E'$, $E''$), S $p_z/p_{xy}$, SOC & 9 & 87 & 26 & 10 & 105 & 61 & 10 \\
(c) Pt & s p d, SOC & 4 & 45 & 28 & 23 & 94 & 23 & 14 \\
(d) Fe & s p d ($t_{2g}$, $e_g$), $\Delta_{t2g}/\Delta_{eg}$, SOC, FM & 4 & 72 & 126 & 15 & 266 & 127 & 15 \\
(e) $\gamma$-Ce & Ce s, d ($t_{2g}$, $e_g$), f ($a_{2u}$, $t_{1u}$, $t_{2u}$) & 3 & 125 & 31 & 9 & 51 & 31 & 9 \\
(f) CeO$_2$ & Ce d ($t_{2g}$, $e_g$), f ($a_{2u}$, $t_{1u}$, $t_{2u}$), O p & 4 & 63 & 53 & 17 & 62 & 33 & 17 \\
(g) MnTe & Mn s d ($A_1$, $E_1$, $E_2$), Te $p_z/p_{xy}$, AFM & 5 & 63 & 72 & 41 & 188 & 124 & 41 \\
EuMn$_2$Bi$_2$ & Eu s d f, Mn s d ($A_1$, $E_1$, $E_2$), Bi s $p_z/p_{xy}$, AFM, SOC on Bi 6p & 13 & 216 & 51 & 32 & 324 & 285 & 32 \\
\bottomrule
\end{tabular*}
\end{table}

One decay law per integral type is too stiff: per-shell integrals reduce the $\pm1$~eV error in every system, from 94 to 23~meV in Pt, 266 to 127~meV in Fe and 105 to 61~meV in MoS$_2$.

Where the crystal field matters, symmetry-split groups cut the error by a further factor of 2--9: MoS$_2$ from 61 to 10~meV, Fe from 127 to 15~meV, $\gamma$-Ce from 31 to 9~meV, CeO$_2$ from 33 to 17~meV, MnTe from 124 to 41~meV and EuMn$_2$Bi$_2$ from 285 to 120~meV (before its added orbitals, Section~\ref{sec:eu}). In Fe the split separates the $t_{2g}$ and $e_g$ exchange ($\Delta=2.30$ and 2.92~eV); in Pt, whose wide d bands need range more than crystal-field freedom, the split model (72 parameters) improves the 4-shell model from 23 to 14~meV.

The Ce f shell split into its cubic groups $a_{2u}/t_{1u}/t_{2u}$ (with d split into $t_{2g}/e_g$) brings $\gamma$-Ce to 9~meV and CeO$_2$ to 17~meV within $\pm1$~eV; the overall RMSE (31 and 53~meV) is dominated by the flat empty f manifold.

SKFitter can also fit non-orthogonal models, with overlap integrals built in the same way as the hoppings (Section~\ref{sec:method}). Graphene uses this option: its sp$^2$ bonding and antibonding bands need the overlaps. The other models are orthogonal.

Six models meet a 25~meV target within $\pm1$~eV (graphene, MoS$_2$, Pt, Fe, $\gamma$-Ce, CeO$_2$), four of them only with symmetry-split groups. The overall RMSE of Fe (126~meV) comes from s--p states 3--8~eV below $E_F$ that carry little weight in the fit.
The decay-law and per-shell models here are fitted from scratch with the same neighbour shells as the final model; they are reference points for the model classes rather than optimised models of their own, and their errors depend on the minimum the fit reaches (with 2 instead of 8 starts the MoS$_2$ decay-law error within $\pm1$~eV is 271 instead of 105~meV).

\subsection{Berry-phase properties of Fe and Pt}\label{sec:berry}

\begin{figure}[!tbp]
\centering
\includegraphics[width=\textwidth]{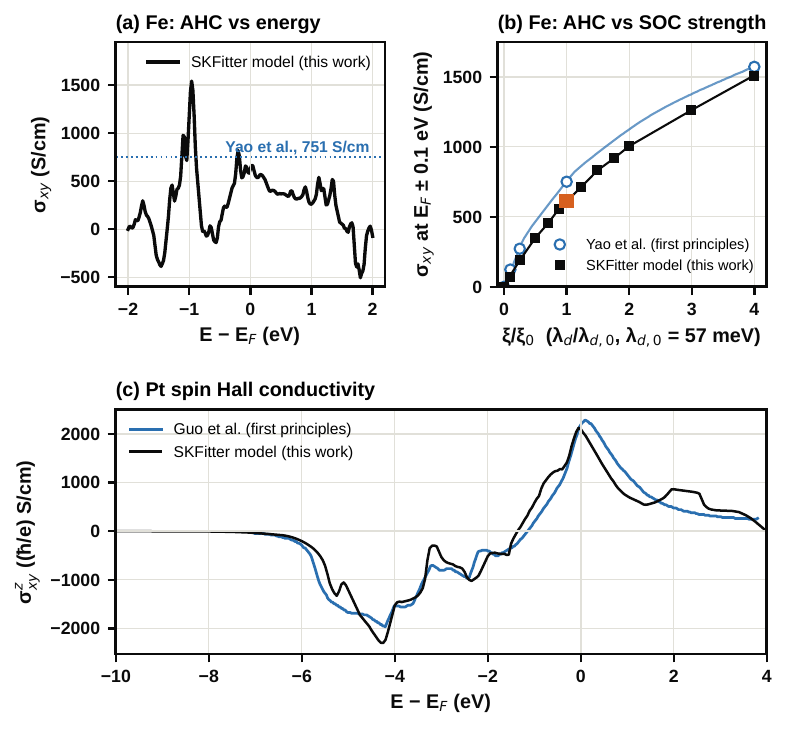}
\caption{\textbf{Berry-phase properties of the fitted models against first-principles calculations.} (a) Anomalous Hall conductivity of bcc Fe against energy ($40^3$ mesh, $k_BT=0.01$~eV); dotted line: first-principles value at $E_F$~\cite{Yao2004}. (b) Fe AHC (mean over $E_F\pm0.1$~eV) against spin--orbit strength relative to its physical value: SKFitter model with $\lambda_d$ scaled from the fitted 57~meV (squares; orange: fitted model) and first-principles values of Yao et al.~\cite{Yao2004} (circles and line, digitized from their Fig.~4). (c) Spin Hall conductivity of Pt from $-10$ to 4~eV: SKFitter s p d model ($80^3$ mesh, black) and first-principles curve of Guo et al.~\cite{Guo2008} (blue, digitized from their Fig.~1b). Fe is plotted as $-\sigma_{xy}$ (Section~\ref{sec:berrymethod}).}
\label{fig:berry}
\end{figure}

We compare the Berry-phase properties of the fitted models with two first-principles references: the anomalous Hall conductivity (AHC) of bcc Fe and its dependence on the spin--orbit strength from a full-potential LAPW calculation~\cite{Yao2004}, and the spin Hall conductivity (SHC) of Pt from a full-potential LMTO calculation~\cite{Guo2008} (Fig.~\ref{fig:berry}). The reference curves were digitized from the published figures (estimated accuracy 2\%; the digitized data are in \code{paper/reference\_data}).

\textbf{Fe anomalous Hall conductivity.} The model gives 628~S/cm at $E_F$ and 612~S/cm averaged over $E_F\pm0.1$~eV, against 751~S/cm in ref.~\cite{Yao2004}. Following ref.~\cite{Yao2004}, we scale the spin--orbit strength $\xi$ relative to its physical value $\xi_0$; in the model this is $\lambda_d/\lambda_{d,0}$ with the fitted $\lambda_{d,0}=57$~meV and all other parameters fixed (one \code{[[property]]} entry per value, Fig.~\ref{fig:berry}b). Both calculations give an AHC that grows sublinearly with $\xi$ over $0\le\xi/\xi_0\le4$: 188, 349, 612, 1004, 1262 and 1510~S/cm at $\xi/\xi_0=0.25$, 0.5, 1, 2, 3 and 4, against 272, 751 and 1572~S/cm at 0.25, 1 and 4 in ref.~\cite{Yao2004}; at $4\xi_0$ the two agree to 4\%. Near $\xi_0$ the model AHC rises by 7.7~S/cm per meV of $\lambda_d$, so $\lambda_d\approx75$~meV would give 751~S/cm. Increasing the fitted spin--orbit parameter improves agreement with the reference calculation, although differences in electronic structure and velocity matrix elements may also contribute to the remaining discrepancy.

\textbf{Pt spin Hall conductivity.} The 4-shell s p d model gives $\sigma^{z}_{xy}=2096$~$(\hbar/e)$~S/cm at $E_F$ on an $80^3$ mesh, against 2200~$(\hbar/e)$~S/cm in ref.~\cite{Guo2008}. Over $-10\le E-E_F\le4$~eV the model follows the reference curve (Fig.~\ref{fig:berry}c): the peak at $E_F$, the decrease above it, the sign change below $E_F$ ($-1.4$ against $-1.1$~eV), the negative extremum ($-2298$ at $-4.25$~eV against $-1970$~$(\hbar/e)$~S/cm at $-4.2$~eV) and the small values below $-6$~eV. Over $-7\le E-E_F\le3.8$~eV, where both curves are available, their correlation is 0.96 and their root-mean-square difference 287~$(\hbar/e)$~S/cm, the largest differences lying 1--3~eV above $E_F$.

\subsection{Altermagnetic spin splitting in MnTe}\label{sec:mnte}

\begin{figure}[!tbp]
\centering
\includegraphics[width=\textwidth]{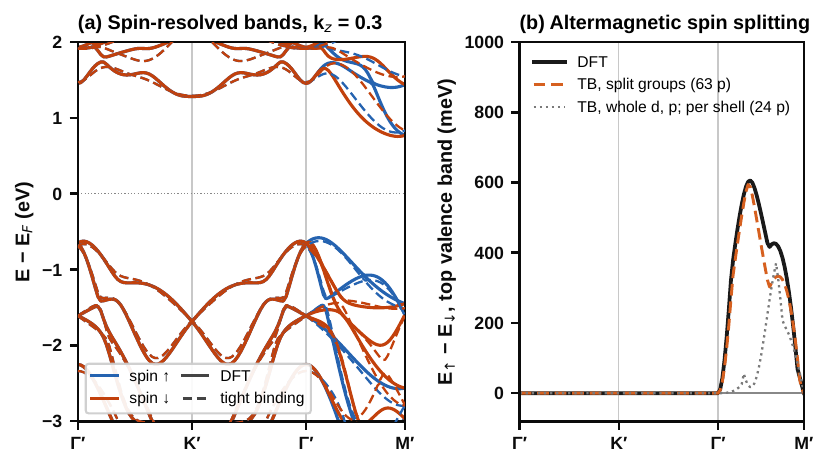}
\caption{\textbf{Altermagnetic spin splitting of MnTe.} (a) Spin-resolved bands at $k_z=0.3$: DFT (solid) and the symmetry-split SK model (5 shells, dashed), spin up (blue) and spin down (orange). (b) Spin splitting $E_\uparrow-E_\downarrow$ of the top valence band: DFT, the symmetry-split model and the per-shell model with whole Mn d and Te p. QE, PBE+U, Mn1~$\uparrow$, Mn2~$\downarrow$, no SOC.}
\label{fig:mnte}
\end{figure}

In NiAs-type MnTe ($P6_3/mmc$) the two Mn sublattices carry opposite moments and are related only by the $6_3$ screw axis, not by a translation or inversion. Their Te octahedra are rotated copies of each other, so spin-up and spin-down bands split away from nodal planes even without SOC~\cite{Smejkal2022PRX031042,Smejkal2022PRX040501,Krempasky2024,Lee2024MnTe}. The bands are taken along $\Gamma'$--$K'$--$\Gamma'$--$M'$ at $k_z=0.3$ (crystal units), a path that crosses both a spin-degenerate and a split direction; \code{skfitter bands -{}-spin} gives the spin-resolved bands (Fig.~\ref{fig:mnte}).

For the top valence band, $E_\uparrow-E_\downarrow$ is zero along $\Gamma'$--$K'$ in both DFT and the model, and reaches 606~meV (DFT) and 594~meV (symmetry-split model) along $\Gamma'$--$M'$, with no altermagnet-specific parameter. The per-shell model with whole Mn d and Te p (124~meV within $\pm1$~eV) has the right sign and nodal directions but reaches only 370~meV: the splitting needs the symmetry-resolved Mn d and Te p hoppings.

The exchange term is the same on both Mn sites up to the sign of $\mathbf n_i$; the splitting arises from the hoppings to the rotated Te environments, as in the symmetry argument. It appears only because the antiparallel site moments are read from the DFT run.

The model (5 shells, 63 parameters) has 72~meV overall and 41~meV within $\pm1$~eV; the per-group exchange splittings are 6.43 ($A_1$), 6.92 ($E_1$) and 7.42~eV ($E_2$).

\subsection{Building up a complex magnet: EuMn$_2$Bi$_2$}\label{sec:eu}

\begin{figure}[!tbp]
\centering
\includegraphics[width=\textwidth]{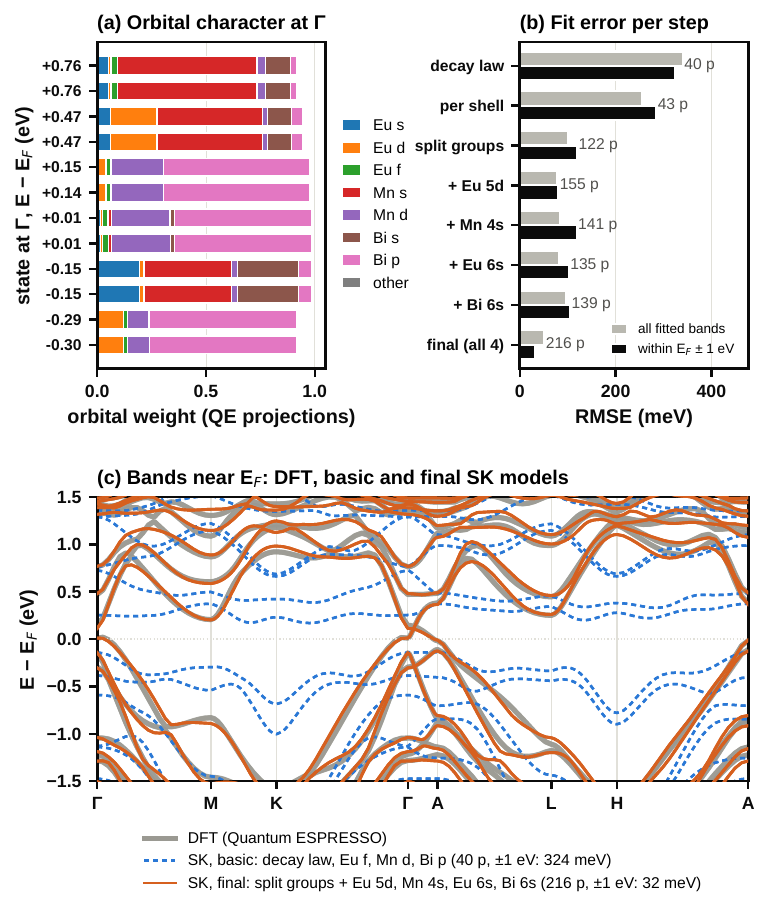}
\caption{\textbf{Building up the EuMn$_2$Bi$_2$ model.} (a) Orbital character of the DFT states at $\Gamma$ within 1~eV of $E_F$ (\code{skfitter character}, Quantum ESPRESSO projections). (b) RMSE over all fitted bands and within $E_F\pm1$~eV at each step: the decay law with Eu f, Mn d and Bi p, per-shell integrals, the symmetry-split groups (Mn $A_1$, $E_1$, $E_2$; Bi $p_z/p_{xy}$), each orbital that \code{skfitter character -{}-suggest} ranks (Table~\ref{tab:suggest}) added alone to the split model, and all four together; numbers are free parameters. SOC on Bi 6p in every step. (c) Bands within 1.5~eV of $E_F$: DFT (grey), the decay-law model (40 parameters, dashed) and the final model (216 parameters, solid).}
\label{fig:eu}
\end{figure}

EuMn$_2$Bi$_2$ combines Eu 4f, Mn 3d and Bi 6p states, exchange on two elements with antiparallel moments (Eu A-type, Mn N\'eel within each layer) and strong Bi SOC in a 10-atom cell. It shows how a model is built up with SKFitter: first by symmetry, then with the orbitals that the DFT projections show to be missing (Fig.~\ref{fig:eu}, Table~\ref{tab:suggest}; \code{examples/eumn2bi2}). SOC acts on Bi 6p only: the Eu data set of the DFT run is scalar-relativistic, and a Mn 3d $\lambda$ fits to zero.

\begin{table}[!tbp]
\caption{The orbitals the symmetry-split EuMn$_2$Bi$_2$ model lacks, as \code{skfitter character -{}-suggest} ranks them over the fitted bands (Section~\ref{sec:suggest}; weights as in the fit), and the RMSE (meV) over all bands and within $E_F\pm1$~eV of the split model with each orbital added alone and with all four. Missing share: the part of the weight the split model cannot carry that the orbital carries; absolute: its largest weight in any state; weighted: its largest band-averaged weight; centre: its projection-weighted energy relative to $E_F$; filled: its occupied weight as a fraction of $2(2\ell+1)$ per site; +param.: parameters added, as predicted by the command (the final model also has the integrals between the added orbitals).}
\label{tab:suggest}
\footnotesize\setlength{\tabcolsep}{3pt}
\begin{tabular*}{\textwidth}{@{\extracolsep\fill}lrrrrrrrr@{}}
\toprule
Model & \hd{missing\\share} & \hd{abs-\\olute} & \hd{weigh-\\ted} & \hd{centre\\(eV)} & filled & \hd{param.} & all & $\pm1$ eV \\
\midrule
split model & & & & & & 122 & 101 & 120 \\
\midrule
+ Eu 5d & 41\% & 0.59 & 0.31 & $+1.6$ & 12\% & +33 & 77 & 81 \\
+ Mn 4s & 22\% & 0.64 & 0.24 & $-1.5$ & 36\% & +19 & 84 & 119 \\
+ Eu 6s & 7\% & 0.25 & 0.06 & $-1.3$ & 26\% & +13 & 82 & 103 \\
+ Bi 6s & 7\% & 0.28 & 0.05 & $-8.9$ & 87\% & +17 & 97 & 105 \\
\midrule
final: all four & & & & & & +94 & 51 & 32 \\
\bottomrule
\end{tabular*}
\end{table}

From one decay law to per-shell integrals to symmetry-split groups, the error within $\pm1$~eV falls from 324 to 285 to 120~meV and over all bands from 341 to 255 to 101~meV (Fig.~\ref{fig:eu}b).

In the split model a steep band pair just above $E_F$ near $\Gamma$ has no counterpart. At $\Gamma$ these states carry Mn 4s, Bi 6s, Eu 5d and Eu 6s weight, orbitals absent from the minimal Eu f, Mn d, Bi p basis (Fig.~\ref{fig:eu}a). This is the band in which first-principles studies place the SOC-driven Bi s--p band inversion at $\Gamma$ that underlies the predicted topological phases of EuMn$_2$Bi$_2$~\cite{Choudhury2024,Sagar2026}.

Over all k-points of the fitted bands, \code{skfitter character -{}-suggest} finds the same four orbitals and orders them (Table~\ref{tab:suggest}): Eu 5d carries 41\% of the weight the split model cannot carry, Mn 4s 22\%, Eu 6s and Bi 6s 7\% each. With the default cutoff the largest weight in any state selects all four and the band-averaged weight Eu 5d and Mn 4s. Bi 6s is 87\% filled, so the models with it count 66 instead of 58 electrons; the predicted parameter counts are exact. For the d shells the command also lists the symmetry groups the $3m$ sites allow ($A_1$, $E_1$, $E_2$), which the model uses for Mn 3d.

Added alone to the split model, Eu 5d gives the largest gain, from 101 / 120 to 77 / 81~meV (all bands / $\pm1$~eV); Eu 6s gives 82 / 103, Mn 4s 84 / 119 and Bi 6s 97 / 105~meV. Over all bands the gains follow the ranking, with Mn 4s and Eu 6s about 2~meV apart; Mn 4s alone barely changes the error near $E_F$.

With all four orbitals (216 parameters) the model reaches 51~meV over all bands and 32~meV within $\pm1$~eV, and the steep band is reproduced (Fig.~\ref{fig:eu}c). The gains are not additive: together the orbitals lower the $\pm1$~eV error by 88~meV, more than the 73~meV of their single gains combined: the states near $E_F$ are mixtures of these orbitals (Fig.~\ref{fig:eu}a).

The Eu f exchange splitting is bounded below ($\geq8$~eV) and the final fit ends at this bound: only the occupied 4f states, 1.8--2.6~eV below $E_F$, lie in the fitted window. Left free, the fit pulls the empty 4f states into the conduction band where DFT has none, a lower RMSE from a wrong model. Likewise the Mn 4s level of the single-orbital fit is kept above $E_F$ (0--8~eV); left free it drops to $-3$ to $-4$~eV and stands in for other missing states.

\section{Comparison with other Slater--Koster codes}\label{sec:compare}

\begin{table}[!tbp]
\caption{Features of SKFitter and other Slater--Koster tight-binding codes, as documented in their papers and manuals. --: not available or not documented.}
\label{tab:compare}
\scriptsize\setlength{\tabcolsep}{3pt}
\begin{tabular*}{\textwidth}{@{\extracolsep\fill}>{\raggedright\arraybackslash}p{\dimexpr0.125\textwidth-2\tabcolsep\relax}>{\raggedright\arraybackslash}p{\dimexpr0.255\textwidth-2\tabcolsep\relax}>{\raggedright\arraybackslash}p{\dimexpr0.122\textwidth-2\tabcolsep\relax}>{\raggedright\arraybackslash}p{\dimexpr0.122\textwidth-2\tabcolsep\relax}>{\raggedright\arraybackslash}p{\dimexpr0.122\textwidth-2\tabcolsep\relax}>{\raggedright\arraybackslash}p{\dimexpr0.122\textwidth-2\tabcolsep\relax}>{\raggedright\arraybackslash}p{\dimexpr0.122\textwidth-2\tabcolsep\relax}}
\toprule
 & SKFitter (this work) & Tight-Binding Studio\newline\cite{Nakhaee2020} & TBFIT\newline\cite{TBFIT} & PySKTB\newline\cite{PySKTB} & tight\-binder\newline\cite{UriaAlvarez2024} & DFTB+\newline\cite{Hourahine2020} \\
\midrule
Purpose & fit SK models to DFT bands; properties & fit SK models to bands (GUI) & fit SK models to bands; topology & build SK models & build and fit SK models; topology, disorder & self-consistent DFTB (energies, forces, transport) \\
Parameters from & DFT bands, read directly (QE XML, VASP vasprun.xml) & bands (VASP XML, OpenMX) & first-principles bands (user-prepared) & user input & bands (user-supplied) & SK tables from atomic DFT \\
Orbitals & s, p, d, f & s, p, d & s, p, d & s, p, d & up to d & s, p, d (f in extended tables) \\
SOC & on-site, every family & on-site & yes & p orbitals only & on-site & on-site \\
Magnetism & exchange per orbital group along each site's moment (collinear and non-collinear), read from DFT & -- & -- & spin-polarised & external fields only & collinear and non-collinear (self-consistent) \\
Symmetry-resolved orbital groups & yes (p, d, f; group-pair integrals, symmetry check, pruning) & -- & -- & -- & -- & -- \\
Distance dependence & per element pair: decay law or independent integrals per shell & per neighbour pair & per neighbour class, scaling & cutoff & neighbour shells; power or exponential law & tabulated from atomic DFT \\
Fitting & orthogonal or non-orthogonal models; bounded least squares or differential evolution, Fermi and symmetry-point weights, warm starts, projection-based state selection and orbital ranking & nonlinear least squares & Levenberg--\allowbreak Marquardt, genetic algorithm, particle swarm & -- & least squares & not band fitting \\
Berry-phase properties & AHC, SHC, valley Berry phases, Berry curvature, Wannier charge centres, $Z_2$, Chern & -- & Berry curvature, Wannier charge centres, $Z_2$, Chern & -- & Berry curvature, Wannier charge centres, $Z_2$, Chern & -- \\
Interface, licence & command line, Python; AGPL-3.0 & GUI, C++; free binary & Fortran; GPL-3.0 & Python; MIT & Python; GPL-3.0 & Fortran, Python API; LGPL-3.0 \\
\bottomrule
\end{tabular*}
\end{table}

Several codes build or fit SK models (Table~\ref{tab:compare}). Tight-Binding Studio~\cite{Nakhaee2020} and TBFIT~\cite{TBFIT} fit SK parameters to band structures, tightbinder~\cite{UriaAlvarez2024} and PySKTB~\cite{PySKTB} build SK models for topology and disorder studies, and DFTB+~\cite{Hourahine2020} uses SK tables computed from atomic DFT for self-consistent total-energy calculations rather than band fitting. SKFitter differs from all of them in four ways:

\textbf{Symmetry-consistent models.} Orbitals can be split into the groups that the site symmetry leaves invariant (p, d and f, including the cubic $a_{2u}/t_{1u}/t_{2u}$ split of f), each pair of groups gets its own integrals, the code checks that every group is invariant under the crystal's operations, and integrals that vanish on every bond are pruned.

\textbf{f electrons with SOC and magnetism in one model.} The two-centre construction covers $ss\sigma$ to $ff\phi$, with on-site SOC on every family and exchange per group along each site's moment direction.

\textbf{Magnetic order from the DFT run.} Moment directions are read from the DFT output, so antiferromagnets, altermagnets and non-collinear orders keep their arrangement (Section~\ref{sec:mnte}).

\textbf{From DFT files to properties.} SKFitter reads Quantum ESPRESSO and VASP output directly, including the orbital projections from which it ranks the orbitals a model lacks (Sections~\ref{sec:suggest} and \ref{sec:eu}) and the starting values, and computes Berry-phase properties from the fitted model, checked here against first-principles results (Section~\ref{sec:berry}).

To summarise, among the implementations considered in Table~\ref{tab:compare}, SKFitter distinguishes itself through the combined treatment of symmetry-resolved orbital-group parameterization, site-dependent magnetic exchange, projection-guided basis refinement, and integrated Berry-phase calculations.
\section{Conclusions}\label{sec:conclusions}

SKFitter turns a DFT band structure into a symmetry-consistent Slater--Koster model with f electrons, SOC and magnetic order read from the calculation, and computes Berry-phase properties of that model. Symmetry-resolved groups add parameters only where the crystal makes them independent, and that is where they pay off: across eight materials they reduce the near-$E_F$ error by factors of 2--9 over per-shell models with whole orbital families. Every parameter is an on-site energy, an exchange splitting, a spin--orbit constant or a two-centre integral between named orbital groups of named elements, so fitted values such as the per-group Mn exchange of MnTe can be compared across materials. The orbitals beyond a minimal basis are chosen from the DFT projections: for EuMn$_2$Bi$_2$ the ranking orders the four missing orbitals by the weight they explain, and together they reduce the near-$E_F$ error from 120 to 32~meV. The models reproduce first-principles Berry-phase results, and properties follow the physics of the model rather than its RMSE: the Fe AHC follows $\lambda_d$ with the same sublinear trend as ref.~\cite{Yao2004}. Large magnetic f--d--p models (216 parameters) take hours (12~h with two threads); analytic gradients and checkpointing are the next steps. 
Further work includes adding the suggested orbitals and groups to a model automatically (the ranking now only prints them), test runs on non-collinear magnets such as Mn$_3$Ir, which the program already supports, and quantum transport with the fitted models. 

\appendix
\section{DFT calculations}\label{app:dft}

All bands were computed with Quantum ESPRESSO 7.5~\cite{Giannozzi2009,Giannozzi2017} and the PBE functional~\cite{PBE1996}, with norm-conserving PseudoDojo~\cite{vanSetten2018} and SSSP~\cite{Prandini2018} pseudopotentials and DFT+U in the simplified rotationally invariant form~\cite{Dudarev1998} (Table~\ref{tab:dft}). Fit bands come from a non-self-consistent run along the high-symmetry path. EuMn$_2$Bi$_2$ is computed in the C-type arrangement (Mn N\'eel within each MnBi layer, layers alike, Eu antiparallel between layers) with moments along $z$; this state is $\mathcal{PT}$ symmetric, so every band is doubly degenerate. First-principles studies place it at or within a few meV per formula unit of the ground state, which depends on U and on the moment direction~\cite{Choudhury2024,Sagar2026}.

\begin{table}[!tbp]
\caption{DFT settings of the band-structure runs (PBE, Quantum ESPRESSO 7.5). FR/SR: fully/scalar-relativistic; NC: norm-conserving; MV: Marzari--Vanderbilt smearing~\cite{Marzari1999}; DFT+U in the Dudarev form~\cite{Dudarev1998}.}
\label{tab:dft}
\scriptsize\setlength{\tabcolsep}{2pt}
\begin{tabular*}{\textwidth}{@{\extracolsep\fill}>{\raggedright\arraybackslash}p{\dimexpr0.224\textwidth-2\tabcolsep\relax}>{\raggedright\arraybackslash}p{\dimexpr0.202\textwidth-2\tabcolsep\relax}>{\raggedright\arraybackslash}p{\dimexpr0.224\textwidth-2\tabcolsep\relax}>{\raggedright\arraybackslash}p{\dimexpr0.135\textwidth-2\tabcolsep\relax}>{\raggedright\arraybackslash}p{\dimexpr0.209\textwidth-2\tabcolsep\relax}}
\toprule
System (structure) & Spin / SOC & Pseudopotentials & DFT+U & Cutoffs, k-mesh, smearing \\
\midrule
Graphene ($a=2.467$~\AA, 20~\AA\ vacuum) & non-magnetic, SOC & pslibrary 1.0.0 FR PAW & -- & 50/400 Ry, $24\times24\times1$, MV 0.005 Ry \\
2H-MoS$_2$ bulk ($a=3.165$, $c=12.440$~\AA) & non-magnetic, SOC & PseudoDojo 0.4 FR NC~\cite{vanSetten2018} & -- & 80/320 Ry, $12\times12\times3$, MV 0.01 Ry \\
Pt fcc ($a=3.943$~\AA) & non-magnetic, SOC & PseudoDojo 0.4 FR NC & -- & 90/360 Ry, $16^3$, MV 0.01 Ry \\
Fe bcc ($a=2.840$~\AA) & non-collinear, SOC, moment along $z$ & PseudoDojo 0.4 FR NC (3s3p semicore) & -- & 100/400 Ry, $16^3$, MV 0.01 Ry \\
$\gamma$-Ce fcc ($a=5.161$~\AA) & non-magnetic, no SOC & SSSP Efficiency PAW~\cite{Prandini2018} (Ce, 4f in valence) & -- & 50/400 Ry, $16^3$, MV 0.02 Ry \\
CeO$_2$ fluorite ($a=5.411$~\AA) & non-magnetic, no SOC & SSSP Efficiency PAW (Ce, O) & -- & 50/400 Ry, $8^3$, Gaussian 0.005 Ry \\
MnTe NiAs-type ($a=4.218$, $c=6.763$~\AA) & collinear, Mn1 $\uparrow$ Mn2 $\downarrow$ ($\pm4.58~\mu_B$), no SOC & PseudoDojo 0.4 SR NC & 4.0 eV (Mn 3d) & 90/360 Ry, $10\times10\times6$, MV 0.01 Ry; bands at $k_z=0.3$ \\
EuMn$_2$Bi$_2$ ($P\bar3m1$, $a=4.802$, $c=15.837$~\AA, cell doubled along $c$) & non-collinear, SOC, Eu A-type, Mn N\'eel in each layer (C-type), moments along $z$ & Eu SSSP PAW (SR); Mn, Bi PseudoDojo 0.4 FR NC & 6.32 eV (Eu 4f), 4.16 eV (Mn 3d) & 85/480 Ry, $8\times8\times2$, MV 0.01 Ry \\
\bottomrule
\end{tabular*}
\end{table}

\section*{CRediT authorship contribution statement}
\textbf{Sundeep Kapila:} Conceptualization, Methodology, Software, Validation, Writing -- original draft. \textbf{Bhaskaran Muralidharan:} Supervision, Funding acquisition, Writing -- review \& editing. \textbf{Satadeep Bhattacharjee:} Supervision, Methodology, Writing -- review \& editing.

\section*{Declaration of competing interest}
SKFitter is distributed under the GNU Affero General Public License v3.0 and may in future also be offered under a commercial licence; it has not been offered or licensed to anyone to date. The authors declare no other known competing financial interests or personal relationships that could have appeared to influence the work reported in this paper.

\section*{Data availability}
The Quantum ESPRESSO data of all calculations, the configurations, the fitted parameters and the scripts that produce every table and figure are part of the SKFitter distribution (\url{https://github.com/sundeep-kapila/skfitter}; data under CC BY 4.0). The release described in this paper is archived on Zenodo~\cite{SKFitterZenodo} (\url{https://doi.org/10.5281/zenodo.23719846}).

\section*{Acknowledgements}
B.M. acknowledges funding from the Department of Science and Technology (DST), Government of India (GoI), under the National Quantum Mission (NQM), through Grant no. DST/QTC/NQM/QMD/2024/4, the Institute of Eminence Grant from IIT Bombay, through Grant no. RD/0523-IOE00I0-139 and the Inani Chair Professorship fund, through Grant No. DO/2024-INAN/001-001. S.B. acknowledges support from KIST (Korea Institute of Science and Technology) Institutional Program: Global Knowledge Platform Research Project (26Z0100). S.K. acknowledges funding from DST, GoI under the NQM and the Swasth Research Fellowship.

\section*{Declaration of generative AI and AI-assisted technologies in the writing process}
During the preparation of this work the authors used Claude (Anthropic) in order to improve language and readability. After using this tool, the authors reviewed and edited the content as needed and take full responsibility for the content of the published article.

\IfFileExists{elsarticle-num.bst}{\bibliographystyle{elsarticle-num}}{\bibliographystyle{unsrtnat}}
\bibliography{references}

\end{document}